\documentclass[aps,prl,twocolumn, superscriptaddress]{revtex4-1}

\usepackage{amsmath,amssymb}
\usepackage{xcolor}
\usepackage{graphicx}
\usepackage{braket}
\usepackage{units}
\usepackage[utf8]{inputenc}

\begin{document}

\newcommand{\h}[1]{\hat{#1}}
\newcommand{\sinc}{\mathrm{sinc}}
\newcommand{\todo}[1]{\textcolor{blue}{ToDo: {#1}}}

\newcommand{\tim}{Applied Physics, University of Paderborn, Warburger Stra\ss e 100, 33098 Paderborn, Germany}

\title{Heralded orthogonalisation of coherent states and their conversion to discrete-variable superpositions}

\author{Regina Kruse}
\email{regina.kruse@upb.de}
\author{Christine Silberhorn}
\author{Tim J. Bartley}
\affiliation{ \tim}

\begin{abstract}
The nonorthogonality of coherent states is a fundamental property which prevents them from being perfectly and deterministically discriminated.  To circumvent this problem, we present an experimentally feasible protocol for the probabilistic orthogonalisation of a pair of coherent states, independent of their amplitude and phase. In contrast to unambiguous state discrimination, successful operation of our protocol is heralded without measuring the states, such that they remain suitable for further manipulation. As such, the resulting orthogonalised state may be used for further processing. Indeed, these states are close approximations of the discrete-variable superposition state $\tfrac{1}{\sqrt{2}}\left(\ket{0}\pm\ket{1}\right)$. This feature, coupled with the non-destructive nature of the operation, is especially useful when considering superpositions of coherent states: such states are mapped to the (weakly squeezed) vacuum or single photon Fock state, depending on the phase of the superposition. Thus this operation may find utility in hybrid continuous-discrete quantum information processing protocols.
\end{abstract}

\maketitle

%
One of the fundamental properties of coherent states is that they are over complete, {i.e.} each coherent state shares some non-zero overlap with every other. 
In the context of state discrimination, this non-zero overlap manifests as errors when one wishes to distinguish two such states. One option to try and discriminate between the two states is by a direct measurement (DM)
. However, as the two states share a finite overlap, we cannot obtain a result with absolute certainty. The limits of the DM approach are determined by a minimal error, the so-called Helstrom bound \cite{helstrom_quantum_1976,kennedy_near-optimum_1972, dolinar_optimum_1973, wittmann_demonstration_2008, sych_practical_2016, muller_robust_2015}. These errors can be overcome using established unambiguous state discrimination protocols~\cite{ivanovic_how_1987}. The original proposals considered schemes in which nonorthogonal states were first orthogonalised, and then measured with a suitable detection scheme~\cite{,dieks_overlap_1988, peres_how_1988}. However, for a valid implementation, these steps must be combined~\cite{dusek_unambiguous_2000, clarke_experimental_2001, roa_quantum-state_2002, barnett_comparison_2003-1, van_loock_implementing_2006, sedlak_unambiguous_2007, jimenez_experimental_2007, roa_conclusive_2010, roa_dissonance_2011, nakahira_optimal_2012, li_assisted_2012, roa_quantum_2014}, such that orthogonalisation is post-selected on the appropriate measurement outcome, non-destructive \cite{bergou_extracting_2013, solis-prosser_experimental_2016} or otherwise. 
Yet, in some cases it may be necessary to orthogalise the input states without a post-selected measurement.

In this paper, we develop a practical scheme for heralded, non-destructive state orthogonalisation of continuous-variable states, namely weak coherent states of opposite phase. This protocol describes the probabilistic transformation of non-orthogonal to orthogonal states, heralded by a predetermined detection event, without destroying the input state. In general, we show how to practically implement the original scheme by Ivanovic and Dieks~\cite{ivanovic_how_1987, dieks_overlap_1988} that aims to transform two non-orthogonal coherent states of the set $\{\ket{\alpha},\ket{-\alpha}\}$ via an operator $\hat{O}$ onto two orthogonal states $\{\ket{\Psi^+},\ket{\Psi^-}\}$, such that $|\bra{\alpha}-\alpha\rangle|>0 \rightarrow |\bra{\alpha}\hat{O}^\dagger \hat{O}\ket{-\alpha}|=|\bra{\Psi^+}\Psi^-\rangle|=0$.
As this procedure is coherent, it may also be used on superposition states $\ket{\Phi_\mathrm{Sup}}=\frac{1}{\sqrt{2}}\left(\ket{\alpha}+e^{i\phi}\ket{-\alpha}\right)$ that are considered the continuous variable equivalent of a qubit~\cite{cochrane_macroscopically_1999, jeong_efficient_2002, ralph_quantum_2003, lund_fault-tolerant_2008}. After transformation with our scheme, the new basis states of the superposition are orthogonal and therefore can be mapped onto the basis states of a discrete variable system. Thus, our scheme finds application as a continuous- to discrete-variable qubit converter.

This paper comprises two sections. We first present the scheme and its application in a state discrimination scenario. The optimal parameters to discriminate the states are found, and we discuss how iterative application of the operation can improve the success probability. In the second section, we look in more detail at the resulting states following this operation. It turns out that these states become extremely good approximations of discrete-variable superposition states. From this, we show how our scheme acts on superpositions of coherent states and returns (weakly squeezed) vacuum and single photon Fock states, depending on the phase of the superposition.
%
%
\begin{figure}
\includegraphics[width=0.65\columnwidth]{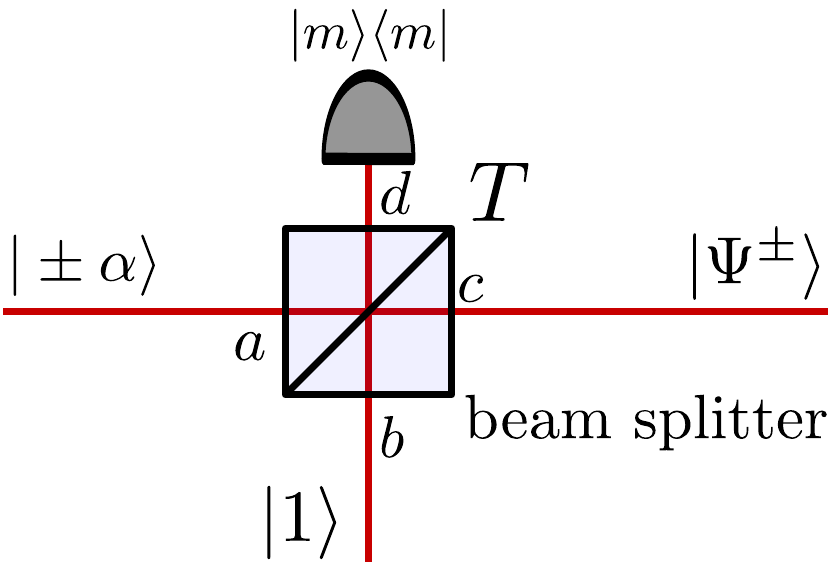}
\caption{(Colour online) Photon replacement scheme for state discrimination. A coherent state from the set $\{\ket{\alpha},\ket{-\alpha}\}$ is incident on one input mode of a beam splitter of predetermined transmissivity $T$. Incident on the other input mode is an ancilla photon. For suitably chosen transmissivity $T$, dependent on $|\alpha|$, a projective measurement of a particular photon number $\ket{m}\bra{m}$ on one output mode transforms the initial set onto the orthogonal set of $\{\ket{\Psi^+},\ket{\Psi^-}\}$ on the other output mode.}
\label{fig:replacement_scheme}
\end{figure}

\section{Heralding non-orthogonal states}
To implement our scheme, we utilise the ``quantum catalysis'' (or ``photon replacement'') technique~\cite{lvovsky_quantum-optical_2002, sanaka_filtering_2006, bartley_multiphoton_2012}, as depicted in Fig.~\ref{fig:replacement_scheme}. An input state from the non-orthogonal set $\{\ket{\alpha},\ket{-\alpha}\}$ is incident on mode $a$ of a beam splitter, simultaneously with an ancilla photon in mode $b$. Dependent on the amplitude of the coherent states $|\alpha|$, we pick the transmissivity of the beam splitter $T\left(|\alpha|\right)$ such that, given a particular outcome of a photon number measurement $\ket{m}\!\bra{m}$ on one output mode of the beam splitter, the input state is projected on either $\ket{\Psi^+}$ or $\ket{\Psi^-}$, depending on the sign of the incident coherent state.
The transformation coefficients of an $n$-photon Fock state for this replacement operation are then given via
\begin{equation}
\begin{aligned}
\ket{\Psi_\mathrm{out}}&=\sqrt{\frac{m!(n+k-m)!}{n!k!}}\sum_{j=0}^{k} \binom{n}{m-j} \binom{k}{j} (-1)^j \\
&\times\sqrt{T}^{n-m+2j}\sqrt{1-T}^{m+k-2j} \ket{n+k-m}_c\otimes \ket{m}_d\, ,
\end{aligned}
\end{equation}
with, in general,  $k$ ancilla and $m$ herald photons for the success event. The transformation of the incident coherent states can then be calculated with the photon number basis representation $\ket{\alpha}=\sum_{n=0}^{\infty}e^{-\frac{\alpha}{2}}\frac{\alpha^n}{\sqrt{n!}}\ket{n}$.

As an example, let us consider the case sketched in Fig.~\ref{fig:replacement_scheme}. At the replacement stage, we define a success event such that one photon is heralded in mode $d$, i.e. $\ket{m}\bra{m}=\ket{1}\bra{1}$. The final output state $\ket{\Psi^\pm}$ is given by
\begin{equation}
\begin{aligned}
\ket{\Psi^\pm}&=\frac{1}{\sqrt{N}}\exp\left(-\frac{|\alpha|^2}{2}\right)\exp\left(\pm\alpha\sqrt{T}\hat{c}^\dagger\right) \\
&\times\left(\pm\alpha(1-T)\hat{c}^\dagger-\sqrt{T}\right)\ket{0}_c\otimes\ket{1}_d
\end{aligned}
\end{equation}
where $\sqrt{\mathcal{N}}$ is the normalisation after the non-unitary transformation and the probability of this event happening is given by $\mathcal{N}=\sum_{n=0}^\infty c_n^2(\mathrm{repl.})$ with $c_n(\mathrm{repl.})=e^{-\frac{\alpha}{2}}\frac{\alpha^n}{\sqrt{n!}}\sqrt{T}^{n-1}[T-n(1-T)]$ as the state coefficients in the photon number basis. A more detailed discussion about the output states can be found in {e.g.} \cite{lvovsky_quantum-optical_2002, sanaka_filtering_2006, bartley_multiphoton_2012}.
\begin{figure}
\includegraphics[width=1.\columnwidth]{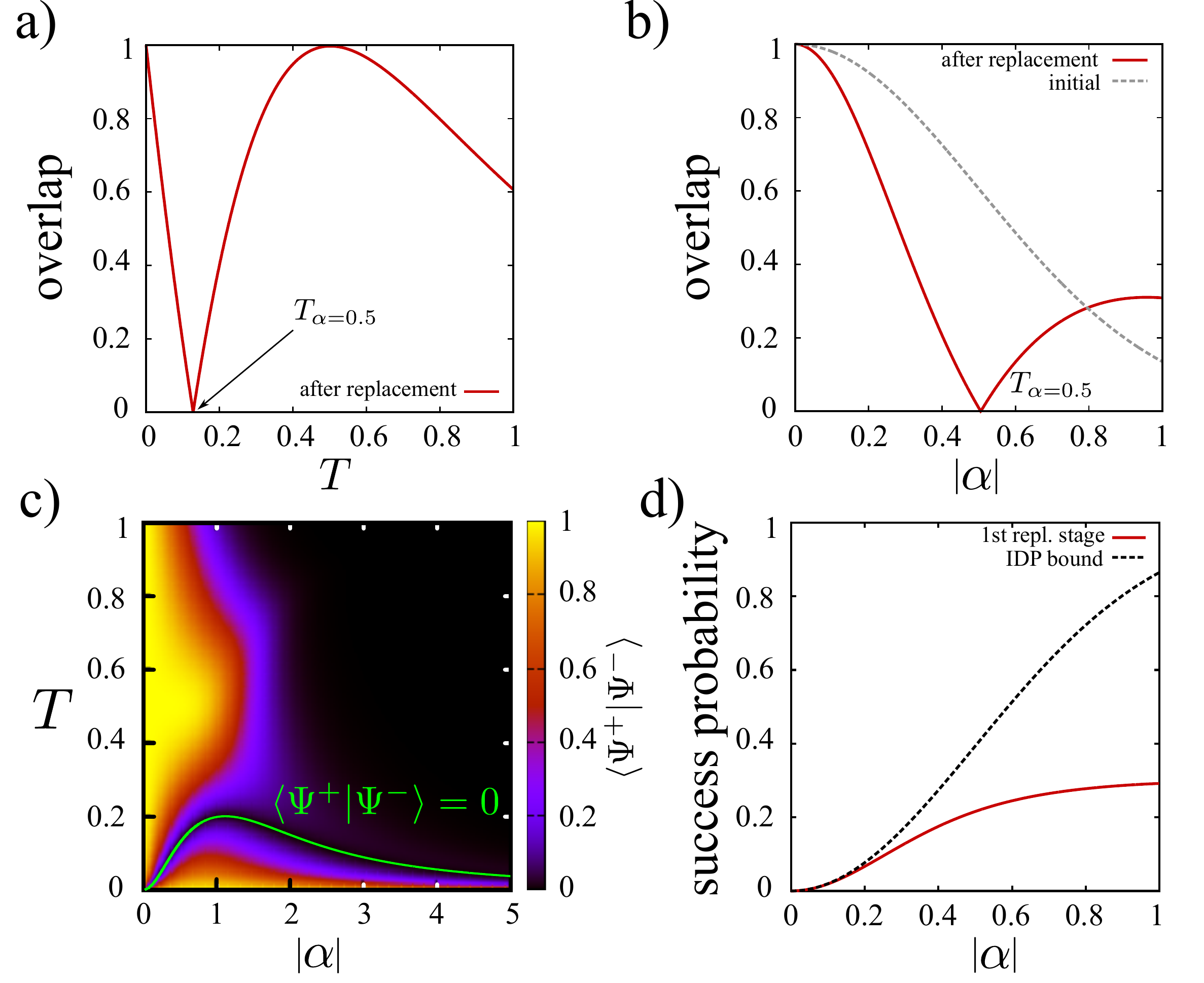}
\caption{(Color online) Overlap after the first replacement stage. Figure (a) shows the overlap after the replacement stage (red) for a fixed amplitude $|\alpha|=0.5$ depending on the beam splitter transmissivity $T$. In figure (b) the overlap after replacement (red) is plotted dependent on $|\alpha|$ for a fixed transmissivity. For comparison the initial overlap is given in grey. In figure (c), we consider both the coherent state amplitude $|\alpha|$ and beam splitter transmissivity $T$ as parameters for the replacement. In green we sketch the line of zero overlap, indicating the optimal beam splitter transmissivity $T(|\alpha|)$ for each $|\alpha|$. Figure (d) depicts the success probability for the first stage of a state discrimination protocol in red compared to the IDP bound in black. For details, see text.}
\label{fig:1st_replacement}
\end{figure}
Writing the output state of the replacement stage in this form illustrates how the state orthogonalisation is possible with this protocol. Calculating the overlap $|\bra{\Psi^+}\Psi^-\rangle|$ and setting it to zero yields a quadratic equation with (at least) one real valued solution due to the different signs of $\alpha$ (a closed-form expression for $T\left(|\alpha|\right)$ is given in the Appendix).

This behaviour is shown in Fig.~\ref{fig:1st_replacement}~(a). In red, we plot the overlap
\begin{equation}
\nonumber
\begin{aligned}
|\bra{\Psi^+}\Psi^-\rangle| &= e^{-(T+1)|\alpha|^2}\\
&\times\left[(1-T)^2T|\alpha|^4-(1-3T)(1-T)|\alpha|^2+T\right]
\end{aligned}
\label{eq:overlap}
\end{equation}
after the replacement stage versus the beam splitter transmissivity $T$. From a transmissivity of $T=0$, we start with an overlap of one, as the beam splitter acts as a mirror and the ancilla photon exits the replacement stage in each case. Increasing the transmissivity, we find that the overlap decreases drastically, before it reaches zero at $T(|\alpha|=0.5)=0.13$. This situation fulfils the aim of the protocol; it is the required beam splitter transmissivity to distinguish states $\ket{+\alpha},~\ket{-\alpha}$ for $|\alpha|=0.5$. Although the shape of the overlap curve is quite steep at this point, this does not pose an unsurmountable experimental challenge since the transmissivity of a beam splitter may be finely controlled.
Going further to $T=1$, the overlap increases to the initial overlap of $|\bra{-\alpha}\alpha\rangle|=0.6$ as the initial states are directly transmitted. In Fig.~\ref{fig:1st_replacement} (b), we consider the overlap after replacement for $T(|\alpha|=0.5)=0.13$ in red. Compared to the initial overlap $|\bra{-\alpha}\alpha\rangle|$ sketched in grey, the overlap of the replaced state drops off faster for small amplitudes $|\alpha|$, before it reaches zero at $|\alpha|=0.5$.

In Fig.~\ref{fig:1st_replacement}(c), we consider the full parameter space for the replacement. We calculate and plot the overlap after replacement depending on the coherent state amplitude $|\alpha|$ and the beam splitter transmissivity $T$. For high amplitudes and splitter transmissivities, we find a large region where the overlap is very small. However, only for the combination of coherent state amplitudes and transmissivities that are represented by the green line the overlap becomes zero. Since this is the goal of the state discrimination, this curve defines the appropriate beam splitter transmissivity $T(|\alpha|)$ for any given $|\alpha|$ in the protocol.
The corresponding probability of success, i.e. the probability of detecting a desired heralding event is plotted in Fig.~\ref{fig:1st_replacement}(d). For each amplitude $|\alpha|$, we calculated the optimal transmissivity $T(|\alpha|)$, where $|\bra{\Psi^+}\Psi^-\rangle|=0$. Further details are provided in the Appendix. From the photon number coefficients $c_n(\mathrm{repl.})$, we have then determined the success probability shown in red. Comparing the success probability to the IDP bound (black) from unambiguous state discrimination for small $|\alpha|$, we find that we already operate close to the optimum predicted for probabilistic discrimination protocols. However, for larger amplitudes we are still some way from optimal operation.

\begin{figure}
\includegraphics[width=0.75\columnwidth]{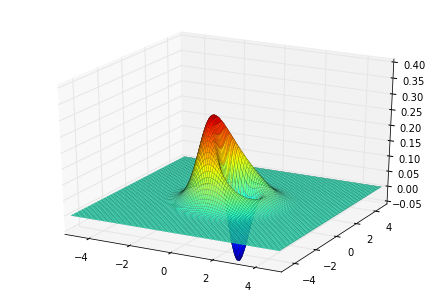}
\caption{(Color online) Wigner function of transformed state $\ket{\psi^+}$. This state has a fidelity $>$98\% with the discrete-variable superposition state $\tfrac{1}{\sqrt{2}}\left(\ket{0}-\ket{1}\right)$ }
\label{fig:WigDisc}
\end{figure}

It is instructive to ask the nature of the state once it has undergone a successful discrimination operation. In fig.~\ref{fig:WigDisc} we plot the Wigner function of the state $\ket{\psi^+}$ when $\alpha=0.5$ and $T=0.13$, the parameters required to discriminate the state from $\ket{\psi^-}$. This state has a fidelity of $>$98\% with the state $\tfrac{1}{\sqrt{2}}\left(\ket{0}-\ket{1}\right)$. Similarly, the state $\ket{\psi^-}$, for which $\alpha=-1/2$ for the same $T$, shares $>$98\% fidelity with the state  $\tfrac{1}{/\sqrt{2}}\left(\ket{0}+\ket{1}\right)$. Thus the discrimination operation heralds the generation of the state $\ket{\alpha}\Rightarrow \tfrac{1}{\sqrt{2}}\left(\ket{0}-\ket{1}\right)$ and $\ket{-\alpha}\Rightarrow \tfrac{1}{\sqrt{2}}\left(\ket{0}+\ket{1}\right)$ with high fidelity. Note that with additional displacement operations, this high fidelity can be achieved for a range of values of $\alpha$.

\begin{figure}
\includegraphics[width=0.95\columnwidth]{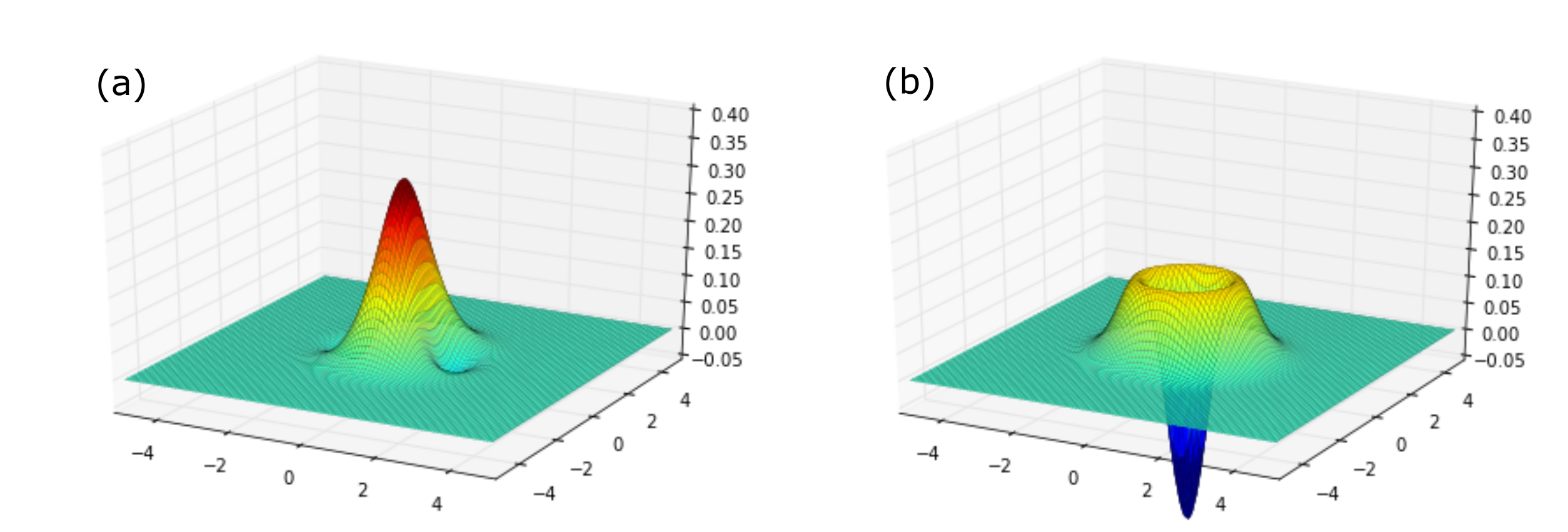}
\caption{(Color online) Wigner functions of (a) transformed state $\ket{\alpha}+\ket{-\alpha}\Rightarrow\ket{\psi^+}-\ket{\psi^-}$ and (b) transformed state $\ket{\alpha}+\ket{-\alpha}\Rightarrow\ket{\psi^+}+\ket{\psi^-}$. These states have $>99\%$ fidelity to the weakly squeezed vacuum and single-photon Fock states, respectively.}
\label{fig:WigCSS}
\end{figure}

We these transformations in mind, we consider the transformations of superpositions of coherent states as inputs. The coherent superposition state $\ket{\alpha}+\ket{-\alpha}$, with amplitude $|\alpha|=0.5$, transforms into the state $\ket{\psi^+}+\ket{\psi^-}$, the Wigner function of which is shown Fig.~\ref{fig:WigCSS} (a). This state has a fidelity of 96\% with the vacuum state $\ket{0}$, but $>$99\% fidelity with the squeezed vacuum state (squeezed by \unit[2.4]{dB}). Similarly the state $\ket{\alpha}-\ket{-\alpha}$ transforms to the state $\ket{\psi^+}-\ket{\psi^-}$, shown in Fig.~\ref{fig:WigCSS}, which exhibits $>$99.9\% fidelity to the single photon Fock state.

\section{Iterative operation}
In contrast to conventional protocols, we have the advantage that our protocol does not destroy the input state, even in case of failure. We can therefore further operate on the state to try to obtain non-orthogonal states. With the information we gain from the heralding event, we can feed forward to subsequent stages. However, due to the large number of possible detection events ($\ket{0}\bra{0},\ket{1}\bra{1},\ket{2}\bra{2}$...) the output states that condition the adaptation of the protocol span a large space. In this paper, we restrict ourselves to a few specific examples and will not exhaust the full parameter space of cascading multiple stages with multiple failure modes.

\begin{figure}
\includegraphics[width=.9\columnwidth]{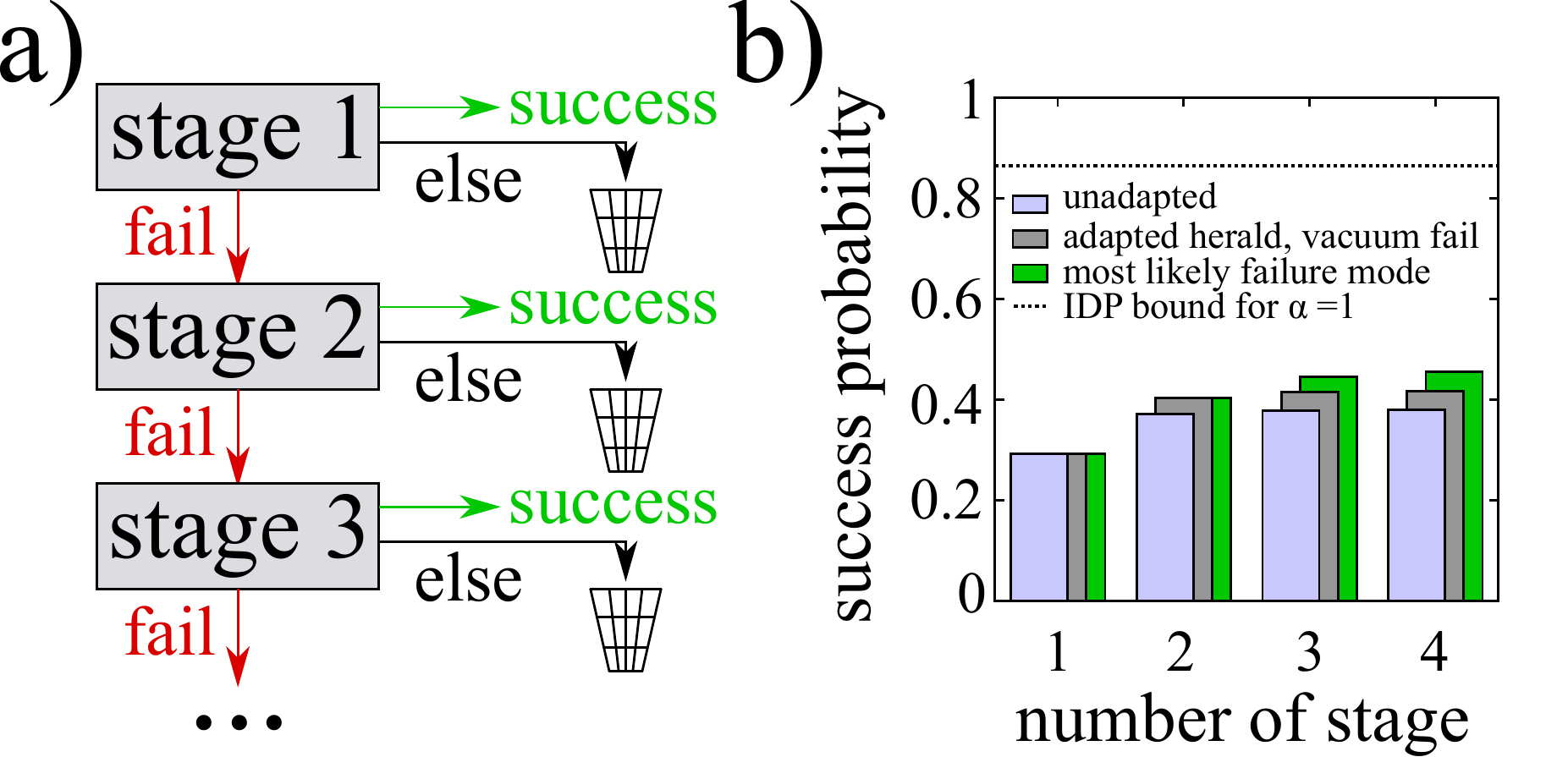}
\caption{(Color online) Cascading of replacement stages. Figure (a) shows the cascading strategy. We consider only one success and one failure heralding event. In any other case, we abort the protocol. In figure (b), we depict the probability of success for three different cascading protocols with $|\alpha|=1.0$. The unadapted protocol (lilac) considers just repeating the first stage with modified transmissivity. The only considered failure mode is vacuum. In grey, we show the success probabilities for a protocol with adapted success events. In this case, only the vacuum is considered as failure mode. The probabilities presented in green are calculated from a protocol that both adapts the success and failure mode to the most likely event at each stage. In all cases, detection outcomes apart from the success and selected failure mode are discarded.}
\label{fig:cascading}
\end{figure}

In particular as sketched in Fig.~\ref{fig:cascading}(a), we only consider two detection events as valid and discard the rest. Let us start with a known $|\alpha|$ at the first replacement stage. Here, we consider heralding on one photon as success and measuring vacuum as failure. Other events, such as detecting 2, 3 or more photons are consequently discarded and the protocol aborted. The failure state from heralding on vacuum is known and we can already prepare the second stage to reattempt to obtain orthogonal states. That is, we can determine the success and failure event of the second stage and adjust the splitter transmissivity accordingly. Going one step further, we then also know the state after the second failure event and can adjust the third stage etc.

In Fig.~\ref{fig:cascading} (b), we consider three of these cascading protocols to optimise the success probability for a given amplitude $|\alpha|=1.0$. A simple example is the repetition of the first replacement stage, {i.e.} using one photon as an ancilla and heralding on one photon for success. As the single failure event, we consider measuring vacuum in mode $d$. However, this simple protocol does not give the highest success probability. We can improve it by adapting the success event to the stage number such that we herald on increasing photon number. This means, in stage one we herald on $\ket{1}\bra{1}$, in stage two on $\ket{2}\bra{2}$ and so on. This approach conserves the photon number between initial and output state if we consider measuring vacuum as the single failure mode \footnote{Strictly speaking, the mean photon number of the input state is changed by the non-unitary operation. However, this approach conserves photon number with respect to the total number of input ancilla photons and output (heralding) measured photons.}. While this protocol gives higher success probability as sketched in grey, it is still possible to improve. In green, we have sketched the success probabilities when also adapting the failure mode from vacuum to measuring one photon less than required for the success event. This is the most likely failure event at any given stage of the protocol. As the success probabilities for the higher stage numbers scale with the failure probabilities in the previous ones, choosing a more likely failure event will increase the overall success efficiency.

However, in each of the depicted protocols in Fig.~\ref{fig:cascading} we do not reach the IDP bound in black. Extrapolating the success probabilities, we assume that the considered cases do not reach the optimal success probability of state discrimination, even in the limit of many stages. However, we have not considered the possibility of other failure modes due to the large parameter space. When considering those, we are likely to improve the overall success probability, however it remains to be seen whether a generalisation of this particular protocol indeed reaches the appropriate bound \cite{dieks_overlap_1988}. 

We also note some interesting behaviour of the overlap between the output states behaves in the cases where we detect a failure event. In general, when a state discrimination protocol fails, the overlap after the failure event will be higher than before \cite{ivanovic_how_1987, chefles_quantum_2000}. In the case of optimal success probability (IDP), a failure will project the two initial states onto the same output state and consequently prohibit any further attempts to distinguish \cite{dieks_overlap_1988, peres_how_1988}. In our case, we have more than one failure mode for which we may consider the overlap after replacement.
\begin{figure}
\includegraphics[width=1.\columnwidth]{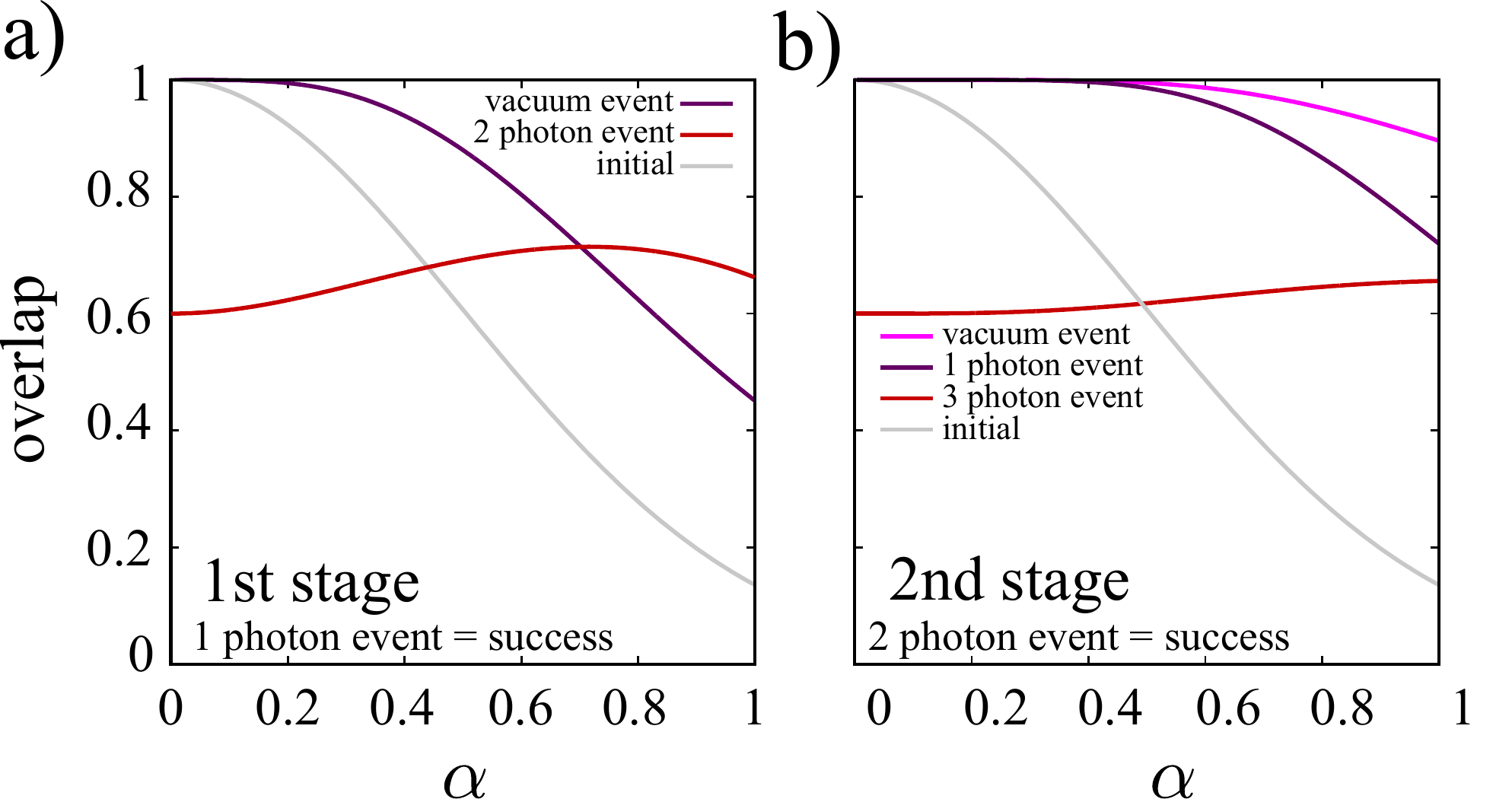}
\caption{(Color online) Overlap of the photon-replaced states when detecting a failure event. In both figures we have plotted the initial overlap between the two stages in grey. Figure (a) shows the overlap after the first stage for the two most likely failure events, i.e. detecting vacuum in purple and heralding on two photons in red. After a vacuum event in the first stage, we also consider the overlap for different failure modes in the second stage in figure (b). For details, see text.}
\label{fig:fail_overlap}
\end{figure}
In Fig.~\ref{fig:fail_overlap}, we plot the overlap for different failure events in the first and second stage compared to the initial overlap. Let us consider the overlap after the first stage in Fig.~\ref{fig:fail_overlap}(a). For comparison, the initial overlap (grey) between the states $\bra{-\alpha}\alpha\rangle$ is depicted dependent on $|\alpha|$. In case of measuring vacuum (shown in purple), the overlap after the replacement increases as expected. Especially in the case of low coherent state amplitudes $|\alpha|\lesssim 0.2$, where the success probability is close to the IDP limit, the overlap after the transformation approaches unity. This behaviour prohibits further distinguishing of the states, as expected.

The interesting case occurs when we detect more photons than necessary for the success event, i.e. more than one in the first stage. Then, for small coherent state amplitudes the overlap \textit{decreases} compared to the initial states. This is atypical for state discrimination protocols and we attribute this effect to the presence of many failure modes where the overlap after failure can be ``distributed'' among them.
To verify that this is not only true for this special case in the first stage, we also compare the overlap after the second replacement stage in Fig.~\ref{fig:fail_overlap}(b). In this case, both the vacuum case (pink) and single photon herald case (purple) show an increase in overlap compared to the initial state (grey). However, when detecting a higher photon number than required for the success event (here: three photons (red)) the protocol behaves atypically and the overlap decreases.

\section{Conclusion}
In conclusion, we have proposed and discussed a scheme for practical heralded state discrimination. We have shown that for each coherent state amplitude $|\alpha|$ we can find a beam splitter transmissivity $T(|\alpha|)$ such that $\{\ket{\alpha},\ket{-\alpha}\}$ is probabilistically mapped onto an orthogonal set $\{\ket{\Psi^+},\ket{\Psi^-}\}$. We have discussed the success probability of such a scheme for cascaded replacement stages and have observed that the overlap after replacement behaves differently to USD for certain failure modes. From these transformations, we have shown that one can construct a heralded coherent converter from continuous-variable coherent states to disccrete-variable superposition states, and from coherent state superpositions to eigenstates of the photon number basis with reasonably high fidelities. This makes this operation a potentially valuable tool in hybrid quantum information processing.

\vspace{0.5cm}
\textbf{Acknowledgements}\par
The authors thank D. Sych for discussions. R.K. received financial support from the European Union’s Horizon 2020 research and innovation program under the QCUMbER project Grant number 665148. T.J.B. acknowledges financial support from the DFG (Deutsche Forschungsgemeinschaft) under SFB/TRR 142.
\bibliography{phot-repl-state-discrimination}
\clearpage
\onecolumngrid
\section{Appendix}
\subsection{Analytical results for first replacement stage}
In the main text, we have claimed that we find at least one real valued solution for the beam splitter transmissivity to set the overlap in equation \eqref{eq:overlap} to zero. While for our case we find three real valued solutions only one is bounded between zero and one. This is the branch that we consider to implement the physical transmissivities
\begin{equation}
\nonumber
\begin{aligned}
T(\alpha)&=\frac{1}{12|\alpha|^4}\left(4|\alpha|^2(3+2|\alpha|^2)-\frac{4(-2)^{1/3}|\alpha|^4(6+|\alpha|^4)}{\left[|\alpha|^6[27-2|\alpha|^2(9+|\alpha|^4)]+3\sqrt{3}\sqrt{-|\alpha|^{12}\{5+4|\alpha|^2(9+|\alpha|^2+|\alpha|^4)\}}\right]^{1/3}}\right.\\
&+\left.2(-2)^{2/3}\left[|\alpha|^6[27-2|\alpha|^2(9+|\alpha|^4)]+3\sqrt{3}\sqrt{-|\alpha|^{12}\{5+4|\alpha|^2(9+|\alpha|^2+|\alpha|^4)\}}\right]^{1/3}\right)\, .
\end{aligned}
\end{equation}
Furthermore, we give a formula to calculate the success probability for the first replacement stage
\begin{equation}
\nonumber
P_\mathrm{success}=e^{-(1-T(\alpha))|\alpha|^2} \left[T(\alpha)+(1-T(\alpha))(1-3T(\alpha))|\alpha|^2+(1-T(\alpha))^2T(\alpha) |\alpha|^4\right]\, .
\end{equation}

\end{document}